\documentclass[twocolumn,showpacs,preprintnumbers,amsmath,amssymb,superscriptaddress]{revtex4}
\usepackage{graphicx}
\usepackage{dcolumn}
\usepackage{bm}
\usepackage{longtable}

\newcommand{\bea}{\begin{eqnarray}}
\newcommand{\eea}{\end{eqnarray}}

\usepackage{amsfonts}
\usepackage{amssymb}
\usepackage{amsmath}
\usepackage{amsthm}
\usepackage{graphicx}

\newcommand{\be}{\begin{equation}}
\newcommand{\ee}{\end{equation}}
\newcommand{\ba}{\begin{eqnarray}}
\newcommand{\ea}{\end{eqnarray}}

\def\Be'{\beta_\mu^{'}}

\def\<{\bigl\langle}
\def\>{\bigr\rangle}

\begin{document}


\title{A stochastic approach for quantifying immigrant integration: the Spanish test case.}

\author{Elena Agliari}
\affiliation{Dipartimento di Fisica, Sapienza Universit\`{a} di Roma, Piazzale Aldo Moro 2, 00185, Roma, Italy}
\affiliation{INdAM, Gruppo Collegato dell'Universit\`a di Roma ``Tor Vergata'', Dipartimento di Matematica, via della Ricerca Scientifica 1, 00133, Roma, Italy}
\author{Adriano Barra}
\affiliation{Dipartimento di Fisica, Sapienza Universit\`{a} di Roma, Piazzale Aldo Moro 2, 00185, Roma, Italy}
\affiliation{INdAM, Gruppo Collegato dell'Universit\`a di Roma ``Tor Vergata'', Dipartimento di Matematica, via della Ricerca Scientifica 1, 00133, Roma, Italy}
\author{Pierluigi Contucci}
\affiliation{Dipartimento di Matematica, Universit\`{a} degli Studi di Bologna, piazza Porta San Donato 2, 00124, Bologna, Italy}
\author{Rickard Sandell}
\affiliation{Departamento de Ciencias Sociales, Universidad de Carlos III de Madrid, 	Avenida de la Universidad 30, 28911, Madrid, Spain}
\author{Cecilia Vernia}
\affiliation{Dipartimento di Matematica, Universit\`{a} degli Studi di Modena e Reggio, Corso Canal Grande 64 01234, Modena, Italy}

\date{\today}

\begin{abstract}
We apply stochastic process theory to the analysis of immigrant integration. Using a unique and detailed data set from Spain, we study the relationship between local immigrant density and two social and two economic immigration quantifiers for the period $1999-2010$. As opposed to the classic time-series approach, by letting immigrant density play the role of ''time", and the quantifier the role of ''space" it become possible to analyze the behavior of the quantifiers by means of continuous time random walks. Two classes of results are obtained. First we show that
social integration quantifiers evolve following pure diffusion law, while the evolution of economic quantifiers exhibit ballistic dynamics. Second we
make predictions of best and worst case scenarios taking into account large local fluctuations. Our stochastic process approach to integration lends itself to interesting forecasting scenarios which, in the hands of policy makers, have the potential to improve political responses to integration problems. For instance, estimating the standard first-passage time and maximum-span walk reveals local differences in integration performance for different immigration scenarios. Thus, by recognizing the importance of local fluctuations around national means, this research constitutes an important tool to assess the impact of immigration phenomena on municipal budgets and to set up solid multi-ethnic  plans at the municipal level as immigration pressure build.
\end{abstract}



 \maketitle

\section{Introduction}

A particular political challenge of growing immigration is immigrant integration. It is considered a necessity for minimizing frictions and confrontation between immigrants and natives in the host community, as well as a precondition for a competitive and sustainable economy\cite{rick1}. In response
to the recent rapid growth in the number of immigrants throughout many major regions in the world, the need for political intervention targeting integration has become increasingly urgent \cite{Castles-2009}. Still, effective policymaking in this area is obstructed by the lack of rudimentary knowledge about how immigrant integration responds to an increase in immigration.

To this end, in a recent work \cite{bcsv} a new approach for studying key-integration quantifiers, based on methods, models, and ideas from statistical physics, was proposed.
The theory describes and predicts how typical integration
quantifiers change when the density of migrants increases. The results predicted a linear growth for the averages of
economic quantifiers like permanent and temporary jobs given to immigrant, and a square root
growth for the averages of social quantifiers like mixed marriages and newborns to mixed couples. This framework is
a powerful tool for the policy makers that are interested in assessing and evaluating integration progresses at the \emph{national level}.

To deal with the phenomena at \emph{municipality level} we use here a different theoretical framework based on the theory and techniques of continuous random walks \cite{Weiss1994,Sokolov2011}. The approach developed in \cite{bcsv}, based on a full micro-macro statistical mechanics theory, revealed in fact a high efficacy to forecast average values. However, since the developed model does't have yet an exact solution, its related phase space picture is not fully disclosed and doesn't cover yet the structure of the fluctuations around the mean values. The random walk approach that we follow here instead, based on a meso-macro stochastic process, has the advantage to allow for a full analytical control of both
mean values and fluctuations.

We consider classical quantifiers of integration such as the fraction of all temporary and permanent labor contracts given to immigrants, the fraction of marriages with spouses of mixed origin (native and immigrant), and the fraction of newborns with parents of mixed origin.
The evolution of these quantifiers versus the percentage of migrants inside the host country is ``locally erratic'', that is, when looked at a fine level of resolution such as the municipality, it can be thought of as a {\it random walk} where the time change is represented by the change of migrant density in the municipality, and the integration quantifier -- playing the role of the space variable -- changes according to suitable probability distributions defining the stochastic process. Instead of obtaining the evolution of averages via statistical mechanics, with this approach the evolution of averages are here the result of averaging over the whole ensemble of municipalities, i.e., averaging over all the random walks.

From a sociological perspective, the evolution of the quantifiers, with respect to the density of immigrants, is, in fact, a random process whose stochasticity may depend on several exogenous factors driving immigration: fluctuations in the ratio between work demand and work request in the host country \cite{Castles-2009}), or ''biases" resulting from (for example) {\em push-pull} factors \cite{Castles-2009} or different types of network induced migration outcomes \cite{Rick2,Rick3,Rick4}. However our aim here is not to explain or disentangle these mechanisms, but rather to look at the evolution of quantifiers as a combined effect of a ''drift" in the presence of some ''noise" regardless of its source/origin. To this task we use random walk theory: the latter constitutes the prototype of stochastic process, and, at the same time, the basic model of diffusion phenomena and non-deterministic motion. Indeed, applications can be found in the study of, for example, transport in disordered media (e.g., \cite{Montroll-1984}), anomalous relaxation in polymer chains (see e.g.,  \cite{Hughes-JStatPhys1982}), financial markets (see e.g., \cite{Bouchaud-2003}), quantitative analysis in sports (see e.g., \cite{Redner-JQuantAnSport2012}).

Using stochastic process theory allows to get a mesoscopic description of the integration quantifiers behavior and to addresses questions such as whether these socio-economic metrics are determined by memory-less stochastic processes or by processes with long-time correlations.
Moreover, this framework allow us to analyze rare events and non-Markovian quantities which are important determinants for planning, in so far they are key tools for quantifying fluctuations.  That is, we aim to provide efficient tools to help assessing the progress (or deficit) in integration as well as to generate strong predictions for extreme case scenarios at lower administrative levels such as \emph{municipalities}, and thereby, through an interplay between statistical mechanics and stochastic processes, we broaden the scope of practical applications of the quantitative theory of immigrant integration as a whole. Typical questions begging an answer are for example: What is the worst/best case scenario in the two integration branches -- social and economic integration -- in a particular municipality if immigrant density changes from say 5 to 7 percent? And how does the effect magnitude of this change compare to the effect magnitude of an equivalent change at the national level, i.e., average change, or in a similar/dissimilar municipality? In other words, through first-passage-time and maximum-span techniques, we obtain estimates for the expected value of immigrant density for which a particular integration quantifier -- say, the share of immigrant workers or the number of mixed marriages -- reaches a given threshold above which new policies, structures, services, facilities etc., have to be made available.

The work is organized as follows: first we describe the database and the procedures for data extraction (Sec. II), then we explain in details the mapping between the evolution of social quantifier and of a random walk (Sec. III and IV) and we report the related results (Sec. V). Finally, we discuss how such outcomes may be exploited to more effectively set up multiethnic plans and immigration policies in general (Sec. VI).

\section{Data description, analysis and elaboration}
Data considered here refer to quarterly observations during the period 1999 to 2010. It is drawn from Spain's Continuous Sample of Employment Histories (the so called Muestra Continua de Vidas Laborales or MCVL) \footnote{It is an administrative data set with longitudinal information for a $4\%$ non-stratified random sample of the population who are affiliated with Spain's Social Security. We use data from the waves $2005$ to $2010$. The residence municipality is only disclosed if the population is larger than $40 000$.} and from the local offices of Vital Records and Statistics across Spain (Registro Civil) \footnote{These data are compounded by the ``National Statistical Agency" (INE). The residence municipality is only disclosed if the population is larger than $10 000$.}. The former provides detailed data on labor contracts, and the latter provides detailed data on spouses and parents to newborns. Information on the municipalities immigration density are drawn from the Municipal population registers  \footnote{More precisely, we use the size of the immigrant population and the native population in each municipality as reported in the 2001 Census as our baseline. Thereafter, based on the information contained in the ``Statistics over residential variation in Spanish municipalities" and statistics on vital events (births and deaths) as elaborated by Spain's ``National Statistical Agency" (INE), we estimate local immigrant densities for different points in time between 1999 and 2010.}  A unique feature of the Spanish data is that all three data sources include also so called undocumented immigrants, that is, immigrants that lack a residence permit. Undocumented immigrants are usually not included in official statistical sources. However, their assimilation within the immigrant population is often significant and excluding them would underestimate the true size of the immigrant population as well as the frequency of the socio-economic events used to measure integration.

Because ``municipality" is the lowest administrative level for which data on density is available, the individual data on mixed events is aggregated to the level of municipality. From these datasets, for each municipality\footnote{Due to data protection, data on mixed marriages and newborns with mixed parents is only available for municipalities with a population larger than $10 000$. In addition, and due to data protection, municipality coding for the labor contract data is only available if the municipality's population exceeds $40 000$. However, about $85\%$ of Spain's immigrants reside in the included municipalities.} we obtain quarterly time series for the following quantities:

\begin{eqnarray} \label{eq:osservabili}
J_p &=& \frac{\# \textrm{permanent contracts to immigrant}}{\# \textrm{permanent contracts}},\\
J_t &=& \frac{\# \textrm{temporary contracts to immigrant}}{\# \textrm{temporary contracts}},\\
M_m &=& \frac{\# \textrm{mixed marriages}}{\# \textrm{marriages}},\\
\label{eq:osservabili4}
B_m &=& \frac{\# \textrm{newborns with mixed parents}}{\# \textrm{newborns}}.
\end{eqnarray}

As explained below, by studying how the quantities in Eqs.~\ref{eq:osservabili}-\ref{eq:osservabili4} vary with the overall fraction of immigrants, we can unveil the growth law determining their evolution and based on this information make previsions.

In order to assess the evolution of the Immigrants-Natives system, a convenient quantity to use as control parameter
is
\begin{equation}
\Gamma =  N_{imm} N_{nat} / N^2= \gamma(1- \gamma),
\end{equation}
where $\gamma = N_{imm}/N$ is the fraction of immigrants. Indeed, $\Gamma$ provides an intensive measure of the cross-links existing among the communities of natives and of immigrants (however, for small values of $\gamma$, $\Gamma \sim \gamma$, hence we can roughly map the percentage of migrants with the time in our bridge). Moreover, differently from other possible choices such as time, using $\Gamma$ avoids any inaccuracy due to seasonality and allows to directly compare municipalities of different sizes (see also \cite{bcsv}).

Complete time series for data on labor contracts involve $\mathcal{M}_J=124$ municipalities and consist of $2976$ data entries over the period
$2005 - 2010$ which is sampled quarterly (i.e. overall $24$ trimesters). Complete series for data on marriages and newborns involve $\mathcal{M}_M=581$ municipalities and consist of
$23240$ data entries spanning the period $1999 - 2008$ which is sampled quarterly (i.e. overall $40$ trimesters).

Thus, for any municipality $i$, we consider five time series: one for $\Gamma$ and one for each observable in Eqs.~\ref{eq:osservabili}-\ref{eq:osservabili4}, hereafter denoted generically as $X^{(i)}$.

As $\Gamma$ varies, each series $X^{(i)}$ determines a ``path'' in the related space and this point process can be looked at as a continuous-time random walk (CTRW) \footnote{The continuous time random walk (CTRW) was introduced by Montroll and Weiss \cite{Montroll-JMP1965}; see also \cite{Weiss1994,Sokolov2011} for recent reviews and SI for a deeper description.}, where the time variable is given by $\Gamma$, while the space variable is given by $X^{(i)}$, see Fig.~$1$. 
This mapping is fully described in the next section.

Finally, in Fig.~$2$ we show the time series for $X^{(i)}$ and $\Gamma^{(i)}$ vs time (in units of trimesters) to highlight the different shape of paths.

 \begin{figure}\label{fig:Walkers}
\includegraphics[width=9cm]{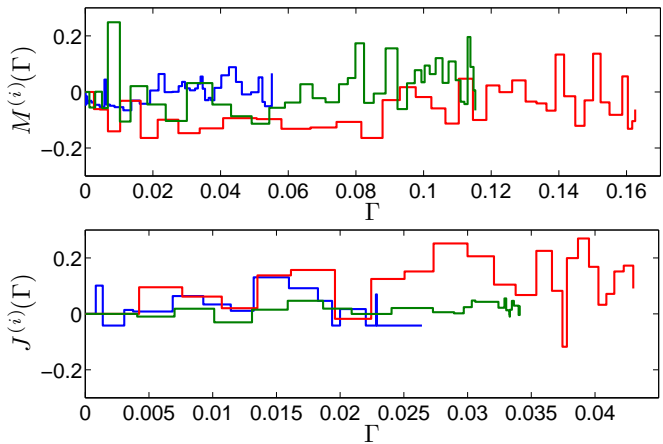}
 \caption{Examples of paths for the quantifiers $M_m$ (upper panel) and $J_p$ (lower panel) shown as a function of $\Gamma$. Three different municipalities are depicted in different colors. These paths can be compared with a theoretical one depicted in Fig.~$3$ and related to a CTRW.}
 \end{figure}

\begin{figure}\label{fig:Walkers_T}
\includegraphics[width=9cm]{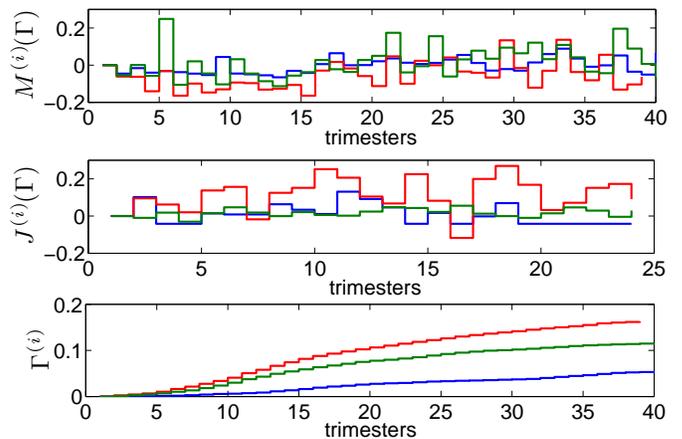}
 \caption{Examples of paths for the quantifiers $M_m$ (upper panel) and $J_p$ (lower panel) shown as a function of time (1 unit = 1 trimester). Three different municipalities (the same as in Fig.~$1$) are depicted in different colors. Notice that for marriages seasonality effects emerge: during summer months marriages are more frequent.}
 \end{figure}

\subsection{Telegraphic introduction on CTRWs}

A CTRW process can be depicted as a dynamical point (to fix ideas embedded in a one-dimensional space, as here we need such a case only), which occupies a position $r(t)$ at time $t$ (see also Fig.~$3$). Let us suppose that the point starts on the origin, that is $r(0)=0$. Then, it stays fixed to its position until time $t_1$, when it jumps to $ \mathbf{\xi}_1$, where it waits until time $t_2 > t_1$, when it jumps to a new location $ \mathbf{\xi}_1 +  \mathbf{\xi}_2$, and so on. The series $\{ t_1, t_2,...\}$ defines the times of jumping events. The times $\tau_1 = t_1-0, \tau_2 = t_2-t_1$ etc. are called waiting times.

The waiting times $\{ \tau_i \}$ and the width of the instantaneous jumps $\{ \xi_i \}$ are continuous random variables extracted from the distribution $\psi (\xi, \tau)$. The latter determines the long-time properties of the walk: a diverging average waiting time typically corresponds to sub-diffusive behaviors, while a diverging variance for jump widths typically corresponds to super-diffusive behaviors.

In particular, for the so-called decoupled continuous random walk (namely where the distribution $\psi( \mathbf{\xi}, \tau)$ factorizes into $\psi( \mathbf{\xi}, \tau)=f(\mathbf{\xi}) \psi(\tau)$), the waiting times and the instantaneous displacements are mutually independent (identically) distributed random variables.

 \begin{figure}\label{fig:Esempio}
\includegraphics[width=9cm]{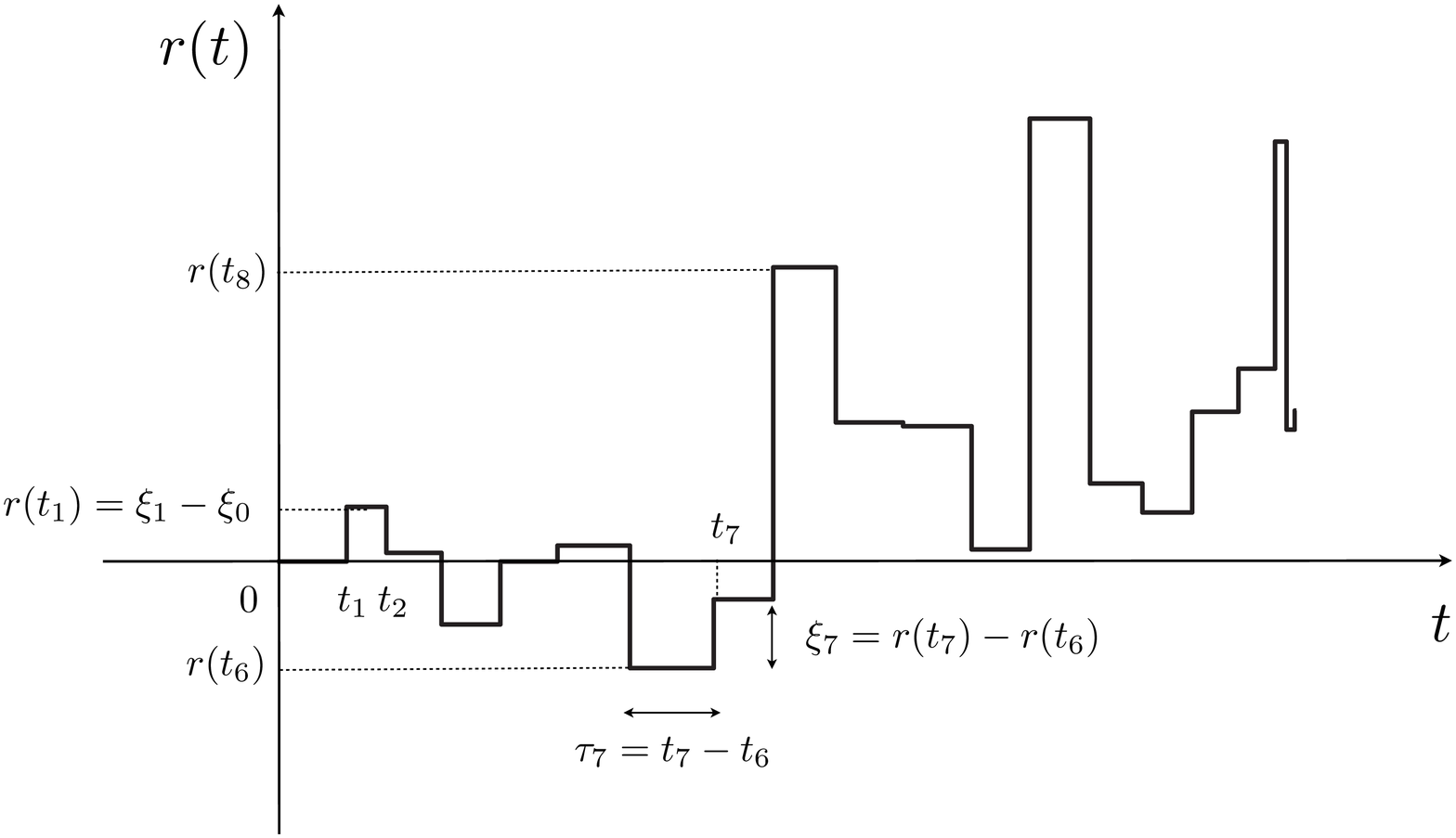}
\caption{Example of path realized by a CTRW whose step widths and waiting times are extracted from the distributions given by Eqs.~\ref{eq:esponenziale} and \ref{eq:normale}, respectively, and with parameters consistent with those found experimentally (see Tab.~II).}
 \end{figure}

The position $r$ of the particle at the $k$-th jump, that is at time $t_k$, is given by the sum $r(t_k) = \sum_{i=1}^{k} \xi_i$. Getting $r(t)$, namely a direct dependence on $t$, requires the introduction of the random variable $n(t)$, representing the number of steps $m$ performed up to time $t$ and defined by $n(t)=\max\{ m: t_m \leq t\}$, in such a way that
\begin{equation}
r(t) = \sum_{i=1}^{n(t)} \xi_i.
\end{equation}

The expected value $\overline{r}(t)$ of the displacement can be derived from the probability distributions for the waiting time and for the step length. In fact, focusing on the decoupled case \footnote{As we will show, this is the case recovered by our experimental data}, we can define $\overline{ \xi}  = \int  \mathbf{\xi}  f(\xi) d \xi $ and $\overline{ \tau}  = \int  \mathbf{\tau}  \psi ( \tau) d \tau$, whereby, as long as $\bar{\tau}$ is finite, one can show that, in the limit of large $t$ \cite{Anomalous}
\be \label{eq:R2b}
\overline{ r (t) } \sim   \overline{ \mathbf{\xi}} \, \frac{t}{\bar{\tau}}.
\ee
Thus, if there is no net drift ($\overline{ \mathbf{\xi}} =0$), the average displacement is zero and one usually looks at the mean square displacement which turns out to scale as $\overline{ r^2 (t) } \sim \overline{ \xi^2} t /  \overline{ \tau}$, and the purely diffusive limit can be recovered.
\newline
On the other hand, in the presence of a net drift ($\overline{ \mathbf{\xi}}  \neq 0$), the mean displacement can also be expressed in terms of the mean number of steps $\overline{ n(t) }$ performed up to time $t$ as (see e.g., \cite{Barkai-PhysA1999,Anomalous})
\be \label{eq:R2}
\overline{ r (t) } = \overline{ n(t) } \cdot  \overline{ \mathbf{\xi}},
\ee
and, accordingly, $\overline{ r^2 (t) } \sim \overline{ r (t) }^2$ \cite{Barkai-PhysA1999,Anomalous}.
From Eq.~\ref{eq:R2}, one can see that if the average time diverges or displays any anomalous behavior, the biased motion turns out to be anomalous as well.


 Of course, the definitions given here can be extended to a geometrical space with arbitrary topology \cite{Weiss1994}.


Despite this random walk process is, by definition, Markovian, one can also introduce non-Markovian related quantities such as the mean-first passage time $\tilde{t}$ and the maximum span $\tilde{r}$, \cite{Majumdar-PhysA2010}.

The mean-first passage time represents the mean time taken by a random walk to first reach a (fixed) point placed at a given initial distance $r$. Its dependence on $r$ qualitatively depends on the kind of diffusion realized, in particular:
\ba
\tilde{t} &\sim& r^2, \,\,\, \textrm{for pure diffusion}\\
\tilde{t} &\sim& r, \,\,\, \textrm{for biased diffusion}.
\ea

The maximum span represents the farthest distance ever reached by a random walk up to time $t$. Again, the functional form of $\tilde{r}$ as a function of $t$ depends on the kind of diffusion realized:
\ba
\tilde{r} &\sim& \sqrt{t} \,\,\, \textrm{for pure diffusion}\\
\tilde{r} &\sim& t,  \,\,\, \textrm{for biased diffusion}.
\ea


These relatively simple laws stem from the peculiarity of the one-dimensioanl structure. In general, the behavior of $\tilde{t}$ and $\tilde{r}$ functionally depends on the underlying topology.

Indeed, due to their non-Markovian nature, estimating such quantities may be rather tricky, yet they are intensively studied as they provide useful information and play an important role in many real situations (e.g. transport in disordered media, neuron firing, spread of diseases and target search processes \cite{Weiss1994,Redner2001,Redner2013}).


To summarize, the CTRW is a stochastic model for which $\psi(\tau)$ and $f( \mathbf{\xi})$ serve as input functions. The output is provided by the temporal series $\{t_1, t_2, ... \}$ and $\{r_1, r_2, ... \}$ from which quantities such as mean squared displacement, mean first-passage time, etc. can be calculated.

In the next section, the jump widths $\xi_i$'s as well as the positions $r(t)$ will assume different meanings (i.e., number of mixed marriages, of newborns from mixed couples, of temporary/permanent contracts to immigrants) according to the
specific quantifier addressed.

\section{The mapping in a nutshell}


Let us denote with $X^{(i)}$ a generic quantifier (i.e., the number of mixed marriages, of newborns from mixed couples, of temporary/permanent contracts to immigrants), where $i$ specifies the municipality.
According to the quantifier considered $i$ is bounded by $\mathcal{M}_J$ or by $\mathcal{M}_M$.

Therefore, we have the time series
\begin{eqnarray}
\{X_1^{(i)}, X_2^{(i)}, ..., X_{\mathcal{T}}^{(i)}\},\\
\{\Gamma_1^{(i)}, \Gamma_2^{(i)},..., \Gamma_{\mathcal{T}}^{(i)}\},
\end{eqnarray}
where $X_n^{(i)}$ and $\Gamma_n^{(i)}$ are the values of the quantifier and of the number of cross-links at the $n$-th trimester and $\mathcal{T}$ is bounded by the overall number of trimesters over which measures have been taken (i.e.,  $24$ for job quantifiers and $40$ for family quantifiers).

For a (one-dimensional) CTRW of $\mathcal{T}$ steps, defined by the two series
\begin{eqnarray}
\{\xi_1, \xi_2, ..., \xi_{\mathcal{T}} \},\\
\{t_1, t_2,..., t_{\mathcal{T}}\},
\end{eqnarray}
where $\xi_n$ is the jump width and $t_n$ is time when the $n$-th step occurs, we recall that the position $r(t)$ of a walker at time $t$ is obtained by $r(t) = \sum_{j=1}^{n(t)} \xi_j$, where $n(t)$ is the number of steps performed up to time $t$.

Analogously, we can state that, for the $i$-th municipality, the value of the quantifier $X^{(i)}(\Gamma)$ corresponding to degree of cross-link $\Gamma$ is
\begin{equation}
X^{(i)}(\Gamma) = \sum_{j=1}^{n^{(i)}(\Gamma)} \Delta X^{(i)}_j,
\end{equation}
where $\Delta X_j^{(i)} = X_{j+1}^{(i)} -X_{j}^{(i)}$ and $n^{(i)}(\Gamma)$ is the latest trimester for which $\Gamma_j^{(i)} < \Gamma$.

Therefore, we can look at the set of $\mathcal{M}$ municipalities as a set of $\mathcal{M}$ random walks.
Actually, before proceeding, a couple of remarks are in order.

In principle, $\Gamma$ and $X$ are bounded by $1$, yet, the number of immigrants corresponds to a small fraction of the overall population in such a way that $\Gamma, X << 1$ and we can neglect boundaries \footnote{Conversely, if boundaries can not be neglected the mapping could still be feasible but we should refer to the theory of random walks on finite chains}.

Moreover, $\Gamma$ and $X$ are not continuous variables as there exists an intrinsic unit given by $1/ \# number \, of \, marriages$, $1/ \# number \, of \, newborns$ and $1/ \# number \, of \, contracts$, representing our experimental sensitivity. However, such a unit is in general much smaller than the quantities measured which can therefore be considered as continuous.

Therefore, we can treat the set of $\mathcal{M}$ municipalities as a set of $\mathcal{M}$ random walks, for which we can build the following ensemble average:
\begin{equation}
\langle X (\Gamma) \rangle \equiv \frac{1}{\mathcal{M}} \sum_{i=1}^{\mathcal{M}} X^{(i)}(\Gamma).
\end{equation}
Similarly, for the average square distance covered
\begin{equation}
\langle X^2 (\Gamma) \rangle \equiv \frac{1}{\mathcal{M}} \sum_{i=1}^{\mathcal{M}} [X^{(i)}(\Gamma)]^2.
\end{equation}

The progression of the quantifiers $\langle X (\Gamma) \rangle$ averaged over the whole set of municipalities, that is to say, the average displacement of the related CTRW, is shown in Fig.~$4$
, where fits evidence the following behaviors
\ba \label{eq:linsqrt1}
\langle J_t(\Gamma) \rangle &\sim& \Gamma,\\
 \label{eq:linsqrt2}
\langle J_p (\Gamma) \rangle &\sim& \Gamma,\\
 \label{eq:linsqrt3}
\langle M_m (\Gamma) \rangle &\sim& \sqrt{\Gamma},\\
 \label{eq:linsqrt4}
\langle B_m (\Gamma) \rangle &\sim& \sqrt{\Gamma}.
\ea
perfectly consistent with those outlined in \cite{bcsv}, despite the procedure for their derivation is conceptually different; this confers robustness to the above results.

To summarize, in our random-walk picture for the time evolution of the social quantifier $X$, in each municipality the quantifier starts from zero and, for a given variation of the related immigrant percentage $\Gamma$, the quantifier increases or decreases until the path ends.
The trajectory of $X$ versus $\Gamma$ qualitatively resembles the position of a CTRW as a function of time (see Figs.~$1$ and $3$
).

In the next section we analyze the CTRWs associated to the quantifiers and try to get a \emph{microscopic} perspective for the origin of these laws. Such a perspective will allow to speculate about possible effects and to make crucial forecasts.

 \begin{figure} \label{fig:SqrtRetta}
\includegraphics[width=9cm]{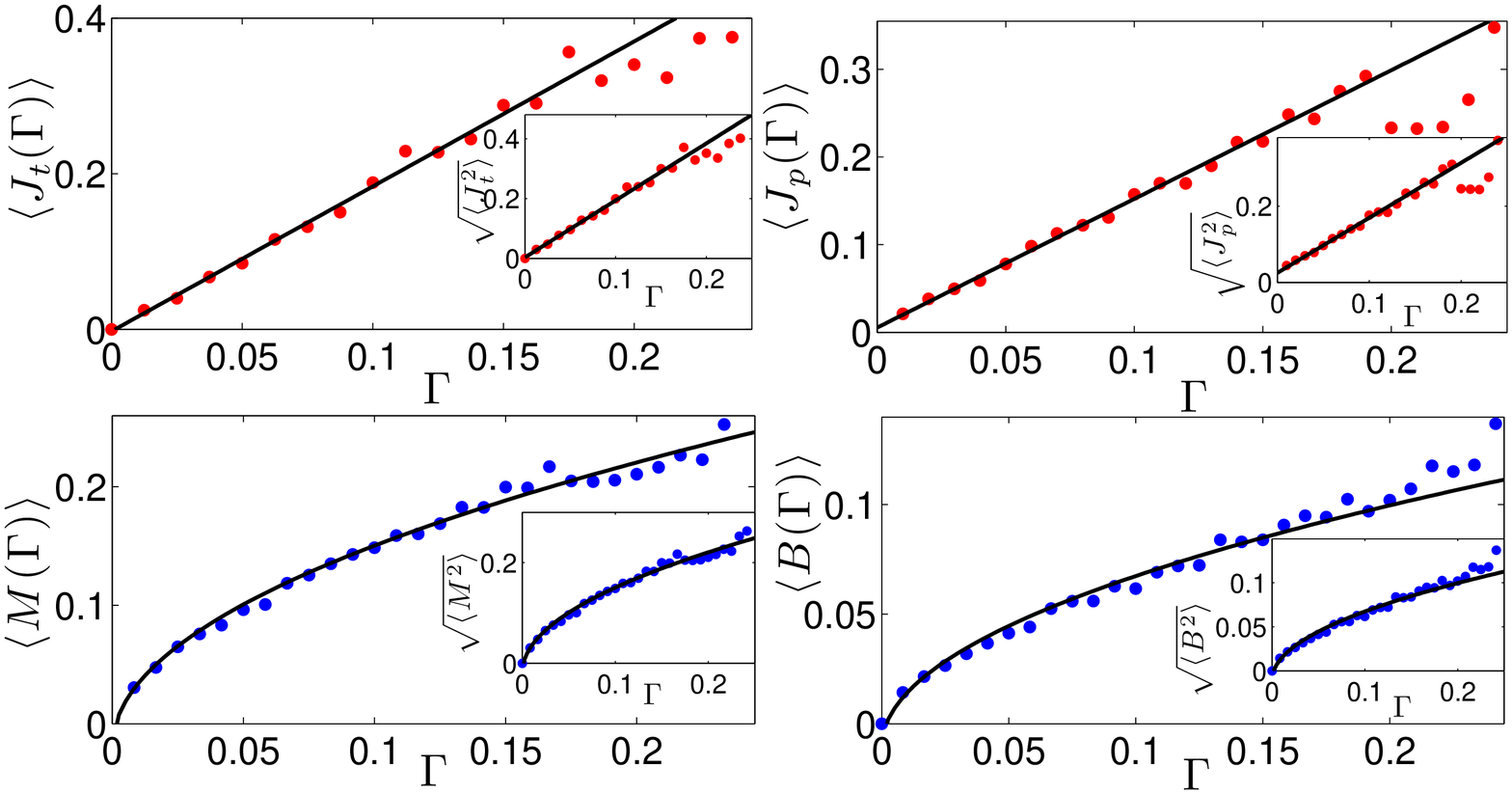}
 \caption{``Mean displacement'' (main figures) and ``mean square displacement'' (insets) versus ``time'' for the CTRWs associated to $J_t$ (panel a), $J_p$ (panel b), $M_m$ (panel c) and $B_m$ (panel d). Data available were binned over $\Gamma$ and averaged over the set of $\mathcal{M}$ municipalities; the resulting values ($\bullet$) and the related best fit (solid line) are shown.
In particular, for family quantifiers we fitted by the law $r=p_1 \sqrt{t}+ p_2$, while for job quantifiers we used the law $r=p_3 t + p_4$; best fit coefficients are summarized in Tab.~I. In general, the goodness-of-fit $R^2$ ranges between $0.97$ and $0.99$.
Notice that $\sqrt{ \langle X^2(\Gamma) \rangle } \sim \langle X(\Gamma) \rangle$ suggests the presence of a drift \cite{Anomalous}.}
\label{fig:SqrtRettaBJt}
 \end{figure}

\begin{table}\label{tab:exp}
\begin{tabular}{|c|c|c|}
\hline
  Quantifier $X$ & $p_1$ & $p_2$ \\
  \hline \hline
  $\langle M_m \rangle$      & $0.54 \pm 0.02$  &    $-0.019 \pm 0.009$ \\
  $\sqrt{\langle M_m^2 \rangle}$      & $0.57 \pm 0.03$  &    $0.007 \pm 0.06$ \\
  $\langle B_m \rangle$    & $0.25 \pm 0.01$   &   $ - 0.010 \pm  0.009$ \\
  $\sqrt{ \langle B_m^2 \rangle} $    & $0.287 \pm 0.002$   &   $- 0.007 \pm 0.004 $ \\
  \hline \hline
  Quantifier $X$ & $p_3$ & $p_4$ \\
  \hline \hline
  $\langle J_t \rangle$  & $1.9 \pm 0.1$  &    $- 0.003 \pm 0.001$ \\
  $\sqrt{\langle J_t^2 \rangle}$  &  $1.9 \pm 0.1$  &    $ 0.003  \pm 0.001$\\
  $\langle J_p \rangle$  & $1.47 \pm 0.06$  &    $0.005 \pm 0.003$ \\
  $\sqrt{\langle J_p^2 \rangle}$  &  $1.45 \pm 0.07$  &    $0.025 \pm 0.008$\\
  \hline
\end{tabular}
\caption{Best-fit coefficient related to plots shown in Fig.~$4$.}
\end{table}



%
%

%




\section{Formalizing the mapping}
\begin{figure}
\includegraphics[width=8cm]{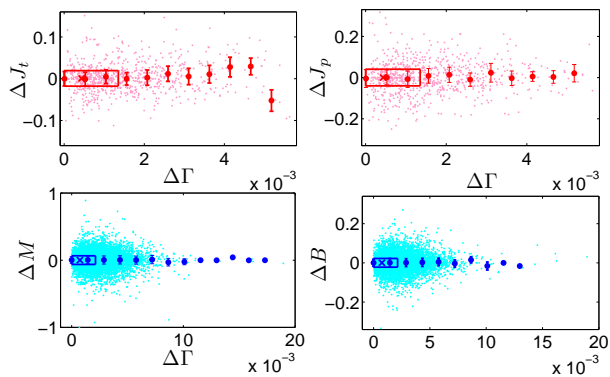}
 \caption{These scatter plots evidence the the existence of any correlation between the ``waiting times'' $\Delta \Gamma$ and the ``jump width'' $\Delta M$ (panel a), $\Delta J$ (panel b), $\Delta J$ (panel c), $\Delta B$ (panel d): each point represents the increments $\Delta X_n$ versus $\Delta \Gamma_n$; all $\mathcal{T}$ steps and the whole set of municipalities are considered. The clouds of data are uniform and do not reveal any special trend. Binned spots evidence the possible values of increments $\Delta \Gamma$ and for each bin we calculated the average of the related increments $\Delta X_n$; the related standard deviation are also depicted. Notice that such averages are basically constant (at least within the error) with respect to $\Delta \Gamma$ and this allows to derive that no clear correlation emerges.
\label{fig:Corr}
}
 \end{figure}

We first check that the CTRWs corresponding to $J_p$, $J_t$, $M_m$ and $B_m$ are decoupled, that is, the related probability distributions $\psi(\Delta X, \Delta \Gamma)$ for the generic increments $\Delta X$ and $\Delta \Gamma$ can be factorized into $f(\Delta X) \psi(\Delta \Gamma)$: this is achieved through direct inspection of the scatter plots reported in Fig.~\ref{fig:Corr}.

%

Thus, we can proceed by studying separately $f(\Delta X)$ and $\psi(\Delta \Gamma)$.
We recall that such distributions provide qualitative information about the diffusive behaviors of the walks associated to our quantifiers, that is, on their time progress.
Moreover, from $f(\Delta X)$ and $\psi(\Delta \Gamma)$, we are able to derive the expectation values
\ba
\overline{\Delta X}  = \int \Delta X  f(\Delta X) d \Delta X,\\
\overline{\Delta \Gamma}  = \int  \Delta \Gamma  \psi(\Delta \Gamma) d \Delta \Gamma,
\ea
which play as the expected jump length and as the expected waiting time respectively. Analogously, we can derive $\overline{ n(\Gamma) }$ which plays as the expected number of steps performed up to ``time'' $\Gamma$, that is
\be
\overline{ n(\Gamma) } = \sum_n n \, Q(n|\Gamma),
\ee
where $Q(n| \Gamma)$ is the probability that $\sum_j^{n} \Delta \Gamma_j$ is smaller than $\Gamma$, but $\sum_j^{n+1} \Delta \Gamma_j$ is larger that $\Gamma$.

From these quantities, one finally has (see e.g., \cite{Barkai-PhysA1999,Anomalous})
\be
\overline{ X (\Gamma) } = \overline{ n(\Gamma) } \cdot  \overline{\Delta X}.
\ee
Of course, the expectation $\overline{ X (\Gamma) }$ and the ensemble average $\langle X (\Gamma) \rangle$ ought to be consistent (as checked in the next section). This ensures the ergodicity of the system and will allow us to exploit the analytical results derived starting from the probability distribution functions also for our ``time'' series.

\subsection{Step width and Waiting time distributions}\label{ssec:time}
Let us start with the distribution for the ``step lengths'' $f(\Delta X)$.
In Fig.~$6$
we show the histogram for the increments $\Delta J_t$, $\Delta M$, $\Delta J_p$ and $\Delta B$ obtained from experimental data.
In all cases the symmetric, centered exponential distribution
\be \label{eq:esponenziale}
f(\Delta X) = \lambda e^{-\lambda |\Delta X|},
\ee
provides an excellent fit.
An exponential distribution for step lengths ensures that  the related CTRW does not exhibit any super-diffusive feature as the central limit theorem is fulfilled.

Now, the fit coefficient $\lambda$ depends on the quantifier considered and it is directly related to the expected value by $\lambda_{X}^{-1} = \overline{\Delta X}$.
Results are collected in Tab.~II
, where a comparison with the experimental average values $\langle | \Delta X|  \rangle$ and $\langle  \Delta X  \rangle$ is also provided.

\begin{table}\label{tab:exp}
\begin{tabular}{|c|c|c|c|}
\hline
  Quantifier $X$ & $\lambda_X^{-1}$ & $\langle |\Delta X| \rangle$ & $\langle \Delta X \rangle$ \\
  \hline \hline
  $J_t$      & $0.031 \pm 0.002$  &    $0.03 $ & $0.003$\\
  $J_p$    & $0.058 \pm 0.003$   &   $0.06 $ & $0.003$\\
  $M_m$  & $0.079 \pm 0.002$  &    $0.08 $ & $0.003$\\
  $B_m$  &  $0.035 \pm 0.001$  &    $0.03 $ & $0.001$\\
  \hline
\end{tabular}
\caption{The second column contains the best-fit coefficients obtained by fitting, according to Eq.~\ref{eq:esponenziale}, the probability distribution function of the displacements $\Delta X$ shown in Fig.~$6$,
 while the third and fourth columns contain the related average values, where the average is performed on raw data over all municipalities.
Being the support of the exponential distribution positive, $\lambda_X^{-1}$ has to be compared with $\langle |\Delta X| \rangle$.
Moreover, we checked that the absolute error on $\langle |\Delta X| \rangle$ is approximately equal to $\langle |\Delta X| \rangle$ itself, as expected from an exponentially-distributed variable. Notice that the average displacement $\langle \Delta X \rangle$ in a single step is positive for any quantifier.}
\end{table}

The goodness of the fit is corroborated by the fact that $\lambda_X^{-1}$ and $\langle | \Delta X|  \rangle$ coincide within the error.
However, looking at $\langle \Delta X \rangle$ we report a slight deviation: while one would expect a null average value due to the centrality of the distribution, the average is systematically positive for all quantifiers and this implies that, as $\Gamma$ increases, $X$ is more likely to grow rather than to decrease. In the random-walk picture, this can be interpreted as the presence of a drift which biases the motion of the walker.


  \begin{figure} \label{fig:HistoDx}
\includegraphics[width=8cm]{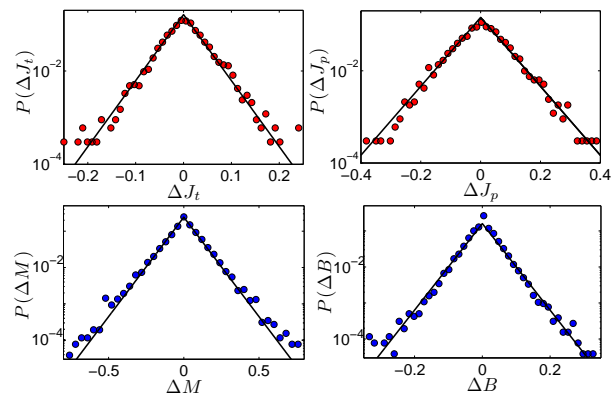}
 \caption{Distributions $f(\Delta M)$ (panel a), $f(\Delta B)$ (panel b), $f(\Delta J_t)$ (panel c) and $f(\Delta J_p)$ (panel d) measured from experimental data, without distinguishing between municipalities, that is, we merged the increments pertaining to the whole ensemble of walks and we built a unique histogram.
Notice the semi-logarithmic scale plot. Data ($\bullet$) are fitted by using Eq.~\ref{eq:esponenziale} (solid line); best-fit coefficients and averages on raw data are collected in Tab.~II
}
 \end{figure}

\quad

\begin{figure}
\includegraphics[width=7cm]{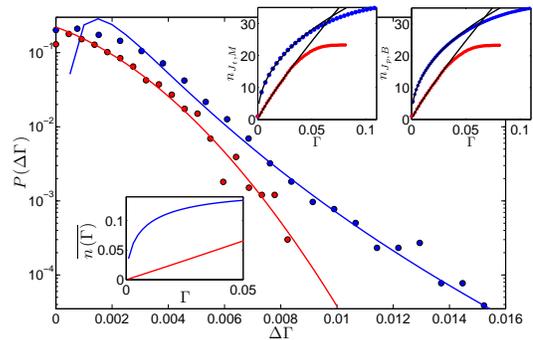}
\caption{Main plot: Histograms for $\Delta \Gamma$, derived from experimental data concerning marriages (blue symbols) and permanent jobs (red symbols), are shown and compared. Solid lines represent the best fit according to a lognormal distribution (see Eq.~\ref{eq:lognormale}) and a half-Gaussian distribution (see Eq.~\ref{eq:normale}), respectively. Fitting coefficients and related errors are reported in Tab.~III
. Notice that such histograms where derived without distinguishing between municipalities.
Lower inset: average number of steps performed up to time $\Gamma$, calculated numerically from Eq.~\ref{eq:lognormale} (red line) and Eq.~\ref{eq:normale} (blue line), respectively.
Upper insets: Average number $\langle n (\Gamma) \rangle$ of steps performed by the related random walker up to the fraction of immigrants $\Gamma$. Solid lines correspond to the law $y \sim x$ and $y \sim \sqrt{x}$, respectively and evidence qualitative different behaviors for marriages and jobs. This picture corroborates the validity of Eq.~\ref{eq:R2} with the ensemble average: $\langle r \rangle \sim \langle n \rangle \langle \Delta r \rangle$, which bridges the picture itself with Fig.~$1$. The fit is robust only up to relatively small values of $\Gamma$, then experimental averages are underestimated. This is due to the fact that the statistics is robust only for values of $\Gamma$ which are reached by (almost) all walks. For larger values our averages are only an underestimate of the expected, effective mean value of $n$.
\label{fig:HistoDT}\label{fig:Steps}}
\end{figure}

Let us now move to the distribution for the ``waiting times'' $\Delta \Gamma$.
\newline
In Fig.~\ref{fig:HistoDT} we show the histogram for the increments $\Delta \Gamma$ obtained from experimental data related to the time period and to the municipalities considered.
%
Interestingly, here qualitative differences emerge between the job quantifiers, i.e. $J_t$ and $J_p$, and the family quantifiers, i.e. $M_m$ and $B_m$.

Before proceeding it is worth stressing that for job quantifiers and family quantifiers the time along which sampling has been performed is not exactly the same, being, respectively, 2005-2010 and 1999-2008 (of course, the consistency between the related time series has been checked for the overlapping period \cite{bcsv}). Now, in order to ensure that the qualitative differences reported do not stem from different time interval, but are intrinsic, we repeated the analysis shown in Fig.~\ref{fig:HistoDT} by restricting only to the common time lapse 2005-2008 and, indeed, we checked the robustness of the result.

In fact, calling $\psi_F$ and $\psi_J$ the distributions for family and job quantifiers respectively, the reason for their intrinsic difference can be depicted in the way mapping between quantifier evolution and random-walks has been fixed. In particular, there exist trimesters $i$ for which a growth in the number of immigrants is reported, i.e. $\Gamma_i - \Gamma_{i-1}>0$,  but no change in the quantifier $X$ considered occurs, i.e. $\Delta X_i =0$. In such cases the two trimesters behave as practically merged as the overall waiting time gets $\Gamma_{i+1} - \Gamma_{i-1}$. This concept can be repeated iteratively until each step of the walk actually corresponds to a true displacement.
Thus, as one can see from Fig.~\ref{fig:HistoDT}, such merging are more frequent for family quantifiers in such a way that the related waiting times  display a larger range. Otherwise stated, the integration of immigrants within the market is more direct: as long as new immigrants arrive, a fraction of them get a job, either permanent of temporary. Conversely, the integration of immigrants from a familiar perspective is more complex and does not follow a prescribed pattern: not surprisingly, the arrival of new immigrants does not necessarily correspond to integration when considering these quantifiers. This is consistent with the results in \cite{bcsv}, where from a different perspective, it is shown that the qualitative difference between the laws $M_m(\Gamma), B_m(\Gamma)$ and $J_t(\Gamma), J_p(\Gamma)$ is due to a different degree of interaction among agents in the two different scenario (families and jobs).

It is worth stressing that such effect is not directly imputable to the seasonality of marriages; this can be seen, for instance, from the fact that for newborns the same effect emerges as well, but their time series do not display any seasonality.

Let us now analyze in more details the waiting time distributions.

For family quantifiers the distribution $\psi_F(\Delta \Gamma)$ fitting the experimental histogram is a log-normal distribution
\be \label{eq:lognormale}
\psi_F(\Delta \Gamma) = \frac{1}{\Delta \Gamma \sqrt{2 \pi} \sigma} \exp{-\frac{(\log \Delta \Gamma - \mu)^2}{2 \sigma^2}},
\ee
for which the average value is expected to be $\overline{\Delta X} = e^{\mu + \sigma^2}$.
As for jobs, the best fit is provided by  a half-normal distribution
\be \label{eq:normale}
\psi_J(\Delta \Gamma) =  \frac{\sqrt{2}}{\sqrt{\pi} \sigma} \exp{-\frac{(\Delta \Gamma - \mu)^2}{2 \sigma^2}},
\ee
for which the average value is expected to be $\overline{\Delta X} = \mu$.
Details on fitting coefficients and average values are all collected in Tab.~III
; notice that, in both cases,  $\overline{\Delta X}$ turns out to be comparable with the ensemble average $\langle \Delta \Gamma \rangle$.

\begin{table*}\label{tab:norm}
\begin{tabular}{|c|c|c|c|c|c|}
\hline
$\Gamma$ & $\mu$ & $\sigma^2$ & $\overline{\Delta \Gamma}$ & $\langle \Delta \Gamma \rangle$ \\
  \hline \hline
  Job  & $(1.2 \pm 0.2) \cdot 10^{-3} $ & $(6.7 \pm 0.6) \cdot 10^{-6} $ & $(2.0 \pm 0.2) \cdot 10^{-3}$ & $ (1.7 \pm 0.2) \cdot 10^{-3}$ \\
  Family & $-6.6 \pm 0.9$ & $0.32 \pm 0.04$ & $(1.7 \pm 0.3) \cdot 10^{-3} $ & $(1.9 \pm 0.3) \cdot 10^{-3}$\\
  \hline
\end{tabular}
\caption{Best-fit coefficients obtained by fitting the probability distribution function of the ``waiting time'' $\Delta \Gamma$ shown in Fig.~\ref{fig:HistoDT} according to Eqs.~\ref{eq:lognormale} and \ref{eq:normale}. The relative error on fit coefficients ranges between $10 \%$ and $20 \%$. Within the error there is perfect consistency between the average values $\overline{\Delta X}$ and $\langle \Delta X \rangle$, as well as between the variance of such distributions and the variance on the related raw data. Here we report only data for marriages and permanent jobs; for newborns and temporary jobs analogous analysis evidence only slight quantitative changes.}
\end{table*}

Thus, although both $\psi_J$ and $\psi_F$ fulfill the central limit theorem and display a finite mean, the latter displays  a long tail so that we expect that the growth for family quantifiers may be slowed down.

In particular, we expect such slowing down to be more evident at ``short times'', namely for small values of $\Gamma$. This can be seen intuitively: for family quantifiers waiting times are more broadly distributed in such a way that for relatively small values of $\Gamma$ it is likely that the the number $n$ of steps performed is rather small, that is, smaller than the mean-field expectation value $\Gamma/\langle \Delta \Gamma \rangle$.

Now, given $\psi_J$ and $\psi_F$, we can derive the number of steps performed up to time $\Gamma$, exploiting the properties of Laplace transforms (see e.g., \cite{Barkai-PhysA1999, Anomalous}).
%
Examples of numerical results of these calculations are shown in the lower inset of Fig.~\ref{fig:HistoDT}: the difference between the two cases is striking.

In order to check this point we measure directly on raw data the average number $\langle n (\Gamma) \rangle$ of steps performed before reaching the time $\Gamma$ (see Fig.~\ref{fig:Steps}).
Indeed, for jobs we find a roughly linear growth, i.e. $\langle n (\Gamma) \rangle \sim \Gamma$,
while for marriages and births we find a slower growth, i.e. $\langle n (\Gamma) \rangle \sim \sqrt{\Gamma}$.


Such a qualitative difference, together with Eq.~\ref{eq:R2}, immediately explains the results of Eqs.~\ref{eq:linsqrt1}-\ref{eq:linsqrt4}.

%

Summarizing, both processes display a non-null positive drift, i.e. $\Delta X >0$, yet the resulting behaviors are qualitatively different over the time window considered. Such a difference ultimately stems from deep differences in the waiting times: a broader distribution for $\Delta \Gamma$ occurs in the case of family quantifiers and the related random walks may experience rather long waiting times, although the jump widths remain narrowly distributed. The net result is just a slowing down in the progress of the quantifier.\\
%

Conversely, as for job, both $\Delta X$ and $\Delta \Gamma$ are narrowly distributed so that at each trimester we do not expect strong variations in the fraction of new immigrants getting a job.

Such a difference suggests an intuitive motivation, namely that the mechanisms underlying the emergence of mixed marriages are more complex and may be subjected to mutual interaction among individuals. This is perfectly consistent with the statistical-mechanics description of the phenomenon provided in \cite{bcsv}.

\section{First predictive outcomes for social planners}

We now turn to the theory's predictive capacity. The aim is to present concrete instruments directed to aid policy makers at the municipal level in their work to accommodate and plan for further immigration. We focus on two well-known observables: the (mean) first passage time, and the (mean) maximum walk span.
\subsection{Mean first-passage time}
Mean first-passage-time quantities have been extensively investigated in a number of different fields, ranging from chemical kinetics to finance, as they provide an estimate for the average time at which a given stochastic event is triggered \cite{Redner2001,Redner2013}.
\newline
Given the process $X(\Gamma)$ we calculate the value $\tilde{\Gamma} (x)$ at which the quantifier reach a certain threshold $x$.
In order to evaluate the typical value of $\tilde{\Gamma} (x)$ we perform an average over the ensemble of walks, that is
\be
\langle \tilde{\Gamma}(x) \rangle = \frac{1}{\mathcal{M}} \sum_{i=1}^{\mathcal{M}} \tilde{\Gamma}^{(i)}(x).
\ee


The quantity $\langle \tilde{\Gamma} (x) \rangle$ allows predictions about the consequences additional immigration have on integration and when a integration threshold is likely to be reached. For instance, let us say that when a integration quantifier reach the threshold  $x$, some integration policies, activities, or services must be activated (e.g. concerning public education, public health, etc). Then, as $\Gamma$ approaches $\langle \tilde{\Gamma}(x) \rangle$ local projects and plans need to be activated.

\begin{figure}
\includegraphics[width=8cm]{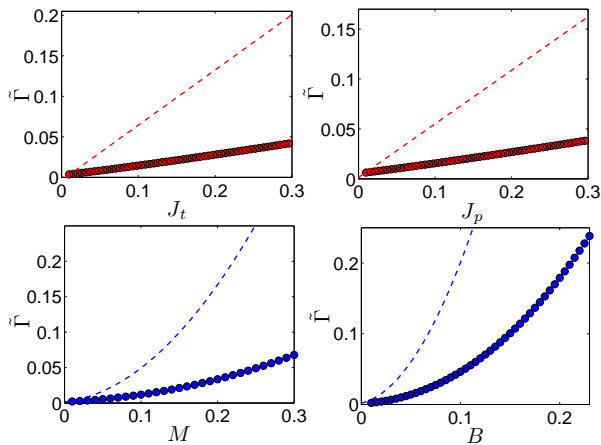}
\caption{Mean time $\tilde{\Gamma}$ to first reach a given value of $J_t$ (panel $a$), of $J_p$ (panel $b$), of $M$ (panel $c$), of $B$ (panel $d$). Experimental data ($\bullet$) are obtained by first getting the mean number of steps to first reach the distance $X$ and then by inverting through $n(\Gamma)$ (see Fig.~\ref{fig:Steps}). Solid lines are best fits given by $y = p_1^{\prime\prime} \Gamma + p_2^{\prime\prime}$ (upper panels) and by $y= p_3^{\prime\prime} \Gamma^2 + p_4^{\prime\prime}$ (lower panels), being $p_1^{\prime\prime}=0.13 \pm 0.02$, $p_2^{\prime\prime}=0.0014 \pm 0.0005$ for $J_p$, $p_1^{\prime\prime}=0.11 \pm 0.02$, $p_2^{\prime\prime}=0.0045 \pm 0.0001$ for $J_t$ and $p_3^{\prime\prime}=0.70 \pm 0.01$, $p_4^{\prime\prime}=0.0047 \pm 0.0003$  for $M_m$, $p_3^{\prime\prime}= 4.54\pm 0.03$, $p_4^{\prime\prime}= 0.00044\pm 0.0002$ for $B_m$. These results are compared with the related $\Gamma (X)$ (dashed line) derived from results shown in Fig.~$4$; see also data in Tab.~I for comparison.}
\label{fig:MFPT}
\end{figure}

In Fig.~\ref{fig:MFPT} we show the mean-first passage time for the quantifiers considered in this work as a function of $X$.
\newline
The mean first-passage time is especially useful for policies plans and service that are coupled with a concrete ``discrete'' integration target, and when we need to know the expected time when the politically defined threshold is reached, and activation of the plans are being called for.

For example, we could ask at which  value of $\Gamma$ (which is the percentage of migrants) we expect that the amount of  newborns from mixed parents reaches the threshold of $10\%$. By simply looking at the behavior of $\langle X (\Gamma) \rangle$, by inverting, we would get $\Gamma \sim 0.2$. However, due to huge fluctuations (hence in some peculiar municipalities), the threshold of $10\%$ can be reached much earlier, as the first passage time, returns a value $\Gamma \sim 0.04$. Hence planning based on average evolutions only may underestimate reality by a factor rendering planning and resource allocation extremely ineffective.


\subsection{Walk span}
The walk span represents the largest point reached by the walker up to a given time. That is, the largest value $\tilde{X}$ reached by $X$ up to $\Gamma$. More precisely, we say that for the $i$-th walk, at the $k$-th step, the span is $\tilde{X}^{(i)}$ if $X(n)^{(i)} < \tilde{X}^{(i)}(k), \forall n \leq k$.
Again, in order to evaluate the typical value of $\tilde{X}(k)$ we perform an average over the ensemble of walks, that is
\be
\langle \tilde{X}(k) \rangle = \frac{1}{\mathcal{M}} \sum_{i=1}^{\mathcal{M}} \tilde{X}^{(i)}(k).
\ee


The average walk span provides information the capacity to integrate further immigration. In fact, in organizing local integration policies and make appropriate priority decisions among different integration initiatives, one is interested in the span of, say, the number of children, or the number of immigrants with permanent jobs, rather than in their average number as the latter may lead to dramatic over- and underestimations.

\begin{figure}
\includegraphics[width=8cm]{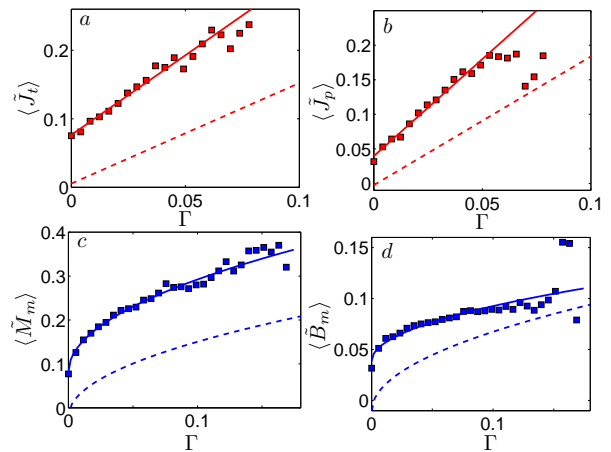}
\caption{Span of the walk for permanent jobs (panel $a$), for temporary jobs (panel $b$), for marriages (panel $c$) and for newborns (panel $d$) versus $\Gamma$. Solid lines are best fits given by $y = p_1^{\prime} \Gamma + p_2^{\prime}$ (upper panels) and by $y= p_3^{\prime} \sqrt{\Gamma} + p_4^{\prime}$ (lower panels), being $p_1^{\prime}=2.3 \pm 0.2$, $p_2^{\prime}=0.08 \pm 0.01$ for $J_p$, $p_1^{\prime}=2.8 \pm 0.2$, $p_2^{\prime}=0.04 \pm 0.01$ for $J_t$ and $p_3^{\prime}=0.8 \pm 0.1$, $p_4^{\prime}=0.04 \pm 0.01$  for $M_m$, $p_3^{\prime}=0.17 \pm 0.02$, $p_4^{\prime}=0.04 \pm 0.01$ for $B_m$.These results are also compared with the curves $X(\Gamma)$ from Fig.~$4$ (dashed line); see also data in Tab.~I for comparison.} \label{fig:Max_Span}
\end{figure}

In Fig.~\ref{fig:Max_Span} we show the span of the quantifiers considered in this work as a function of $\Gamma$.
We notice that the qualitative differences already evidenced for $\langle X(\Gamma) \rangle $ are robust and, the span for marriages and births grows like $\sqrt{\Gamma}$, while the span for temporary and permanent jobs grows like $\Gamma$. The persistence of such behaviors is consistent with the fact that such random walks display distributions for waiting time and step width having finite average and variance. For instance, for a simple random walk on a line the span grows in time like $\sqrt{t}$, while in the presence of a drift one has a linear law $t$ \cite{Weiss1994}.

\section{Conclusions}
Theoretical models, originally developed to solve physical problems, are increasingly being used to study social phenomena.
Statistical mechanics and stochastic process theory are particularly well suited for this task, and have generated a novel
quantitative understanding of the underlying complexity of social interactions. In this paper we focused on stochastic processes.
We identified the random behavior of the four integration quantifiers with random walkers: each municipality draws a random walk in the quantifier-migrant's density plane. Averaging over all the municipalities then allowed to investigate the evolution of the quantifier averages, which are found to scale with the square root of the percentage of migrants for familiar quantifiers and linearly with the percentage of migrants for job quantifiers, in complete agreement with previous findings obtained through the statistical mechanical route $[2]$.
We inferred the distributions of jumps and waiting times (which are found to be decoupled):  while jump distributions are exponentially distributed for all the quantifiers, waiting time distributions depends on the context: social quantifiers have log-normal distributions for those times, while economic quantifiers display Gaussian distributions.

This difference has a simple explanation. While there is a correlation, even on a short timescale between the last-arrived migrant and that migrants incorporation into the labor market (in order to sustain), the same is not true for marriages or newborns. Clearly correlation is likely to be negligible between the last arrived migrant and a mixed marriage or birth event (i.e., it is unlikely that the arriving migrant and the one, say,  marrying a native are the same person). This results in a stronger noise affecting our social quantifiers, which destroys the net drift and simple diffusion is the only survivor. On the contrary, driven by the necessity to work to survive, our economic quantifier display ballistic motion. Another motivation that contributes to the macroscopic differences resides in the much broader distribution of jumps for the working quantifiers: The fat tail encoding for the long jumps in the working quantifiers implies a larger value of drift,  that, coupled with much less noise -- for the reasons just mentioned -- lead to ballistic motion.
\newline

From a practical purpose, no power-law distributions are found. Hence, the Central Limit Theorem holds implying that the theory is suitable for generating predictions. To this end, we introduced two predictive non-Markovian tools: the ''mean first passage time" and the ''maximum span walk". Using these tools it become possible to tackle questions that traditionally been answered using guesstimates in a more scientific way. For example, our predictive framework can easily produce forecast of the share of newborns with mixed parents following an increase in the share of immigrants from, say, 3 to 5 percent?
We make two types of forecasts: first, we assess the evolution of the mean of this quantifier. The evolution is obtained evaluating from Figure $4$ the average increment, which is roughly from $B(\Gamma)= 0.04$ to $B(\Gamma)=0.05$.
Second we assess the mean worst case by dealing with fluctuations. These fluctuations are obtained by extrapolating data from Figure 8, which
gives a $\tilde{B}(\Gamma) \sim 0.08$, i.e. more than fifty percent higher than its average value.
\newline
Although the investigated quantities are non-Markovian ($\langle \tilde{X}(\Gamma) \rangle$ and   $\langle \tilde{\Gamma}(X) \rangle$)
their behavior is still treatable: each of them can indeed be studied separately as a
one-dimensional random walk also concerning the first passage time and the maximum span walk.
\newline
On a broader level, this work provides a concrete rigorous method for quantitative studies of social-science problems. The choice of
immigrant integration is motivated by its prominent place in both the UE and the US political agendas. By uncovering the local variation pattern in the quantifiers we produced a scientific tool for anticipating the consequences of further immigration on local integration process. Information of this type has not been available in the past and constitutes great value for the development of immigration policies and multi-ethnic planning at the local level. However, while this work advances our knowledge on integration phenomena, other effects, like segregation phenomena, that may spontaneously develop in the host country has yet to be considered and incorporated into the theoretical framework developed here.
\section*{Acknowledgments}
This work is supported by the FIRB grants RBFR08EKEV and RBFR10N90W.
\newline
EA and AB acknowledge also partial financial support by GNFM-(INdAM).
\newline
RS is grateful to the project "Competition, Adaptation and Labour-Market Attainment of International Migrants in Europe (CALMA)" granted by the VI National Plan for Scientific Research, Spanish Ministry of Economy and Competitiveness (CSO2012-38521), for partial financial support.

\bibliography{social}

\begin{thebibliography}{18}
\expandafter\ifx\csname natexlab\endcsname\relax\def\natexlab#1{#1}\fi
\expandafter\ifx\csname bibnamefont\endcsname\relax
  \def\bibnamefont#1{#1}\fi
\expandafter\ifx\csname bibfnamefont\endcsname\relax
  \def\bibfnamefont#1{#1}\fi
\expandafter\ifx\csname citenamefont\endcsname\relax
  \def\citenamefont#1{#1}\fi
\expandafter\ifx\csname url\endcsname\relax
  \def\url#1{\texttt{#1}}\fi
\expandafter\ifx\csname urlprefix\endcsname\relax\def\urlprefix{URL }\fi
\providecommand{\bibinfo}[2]{#2}
\providecommand{\eprint}[2][]{\url{#2}}

\bibitem[{ric(2010)}]{rick1}
\emph{\bibinfo{title}{European Commission: Handbook on Integration for
  policy-makers and practitioners}} (\bibinfo{publisher}{Publications Office of
  the European Union}, \bibinfo{address}{Luxembourg}, \bibinfo{year}{2010}).

\bibitem[{\citenamefont{Castles and Miller}(2009)}]{Castles-2009}
\bibinfo{author}{\bibfnamefont{S.}~\bibnamefont{Castles}} \bibnamefont{and}
  \bibinfo{author}{\bibfnamefont{M.~J.} \bibnamefont{Miller}},
  \emph{\bibinfo{title}{The {A}ge of {M}igration - {I}nternational {P}opulation
  {M}ovements in the {M}odern {W}orld}} (\bibinfo{publisher}{Pallgrave
  McMillian}, \bibinfo{address}{{N}ew {Y}ork}, \bibinfo{year}{2009}).

\bibitem[{\citenamefont{Barra et~al.}(2014)\citenamefont{Barra, Contucci,
  Sandell, and Vernia}}]{bcsv}
\bibinfo{author}{\bibfnamefont{A.}~\bibnamefont{Barra}},
  \bibinfo{author}{\bibfnamefont{P.}~\bibnamefont{Contucci}},
  \bibinfo{author}{\bibfnamefont{R.}~\bibnamefont{Sandell}}, \bibnamefont{and}
  \bibinfo{author}{\bibfnamefont{C.}~\bibnamefont{Vernia}},
  \bibinfo{journal}{Sci. Rep. Nature} \textbf{\bibinfo{volume}{4}},
  \bibinfo{pages}{4174} (\bibinfo{year}{2014}).

\bibitem[{\citenamefont{Massey and Zenteno}(1999)}]{Rick2}
\bibinfo{author}{\bibfnamefont{D.}~\bibnamefont{Massey}} \bibnamefont{and}
  \bibinfo{author}{\bibfnamefont{R.}~\bibnamefont{Zenteno}},
  \bibinfo{journal}{{P}roc. \ {N}atl. \ {A}cad. \ {S}ci.}
  \textbf{\bibinfo{volume}{96}}, \bibinfo{pages}{5328} (\bibinfo{year}{1999}).

\bibitem[{\citenamefont{Wilson and Portes}(1980)}]{Rick3}
\bibinfo{author}{\bibfnamefont{K.~L.} \bibnamefont{Wilson}} \bibnamefont{and}
  \bibinfo{author}{\bibfnamefont{A.}~\bibnamefont{Portes}},
  \bibinfo{journal}{American Journal of Sociology}
  \textbf{\bibinfo{volume}{86}}, \bibinfo{pages}{295} (\bibinfo{year}{1980}).

\bibitem[{\citenamefont{Sandell}(2012)}]{Rick4}
\bibinfo{author}{\bibfnamefont{R.}~\bibnamefont{Sandell}},
  \bibinfo{journal}{Int. Migrat. Rev.} \textbf{\bibinfo{volume}{49}},
  \bibinfo{pages}{971} (\bibinfo{year}{2012}).

\bibitem[{\citenamefont{Montroll and Schlesinger}(1984)}]{Montroll-1984}
\bibinfo{author}{\bibfnamefont{E.}~\bibnamefont{Montroll}} \bibnamefont{and}
  \bibinfo{author}{\bibfnamefont{M.}~\bibnamefont{Schlesinger}},
  \emph{\bibinfo{title}{Nonequilibrium Phenomena II: From Stochastics to
  Hydrodynamics}} (\bibinfo{publisher}{North-Holland}, \bibinfo{year}{1984}).

\bibitem[{\citenamefont{Hughes et~al.}(1982)\citenamefont{Hughes, Montroll, and
  Schlesinger}}]{Hughes-JStatPhys1982}
\bibinfo{author}{\bibfnamefont{B.}~\bibnamefont{Hughes}},
  \bibinfo{author}{\bibfnamefont{E.}~\bibnamefont{Montroll}}, \bibnamefont{and}
  \bibinfo{author}{\bibfnamefont{M.}~\bibnamefont{Schlesinger}},
  \bibinfo{journal}{{J}. \ {S}tat. \ {P}hys.} \textbf{\bibinfo{volume}{28}},
  \bibinfo{pages}{111} (\bibinfo{year}{1982}).

\bibitem[{\citenamefont{Bouchaud and Potters}(2003)}]{Bouchaud-2003}
\bibinfo{author}{\bibfnamefont{J.}~\bibnamefont{Bouchaud}} \bibnamefont{and}
  \bibinfo{author}{\bibfnamefont{M.}~\bibnamefont{Potters}},
  \emph{\bibinfo{title}{Theory of Financial Risk and Derivative Pricing}}
  (\bibinfo{publisher}{Cambridge University Press}, \bibinfo{year}{2003}).

\bibitem[{\citenamefont{Gabel and Redner}(2012)}]{Redner-JQuantAnSport2012}
\bibinfo{author}{\bibfnamefont{A.}~\bibnamefont{Gabel}} \bibnamefont{and}
  \bibinfo{author}{\bibfnamefont{S.}~\bibnamefont{Redner}},
  \bibinfo{journal}{Journal of Quantitative Analysis in Sports}
  \textbf{\bibinfo{volume}{1416}} (\bibinfo{year}{2012}).

\bibitem[{\citenamefont{Klages et~al.}(2007)\citenamefont{Klages, Radons, and
  Sokolov}}]{Anomalous}
\bibinfo{editor}{\bibfnamefont{R.}~\bibnamefont{Klages}},
  \bibinfo{editor}{\bibfnamefont{G.}~\bibnamefont{Radons}}, \bibnamefont{and}
  \bibinfo{editor}{\bibfnamefont{I.~M.} \bibnamefont{Sokolov}}, eds.,
  \emph{\bibinfo{title}{Anomalous Transport: Foundations and Applications}}
  (\bibinfo{publisher}{Wiley-VCH}, \bibinfo{year}{2007}).

\bibitem[{\citenamefont{R.~Metzler and Klafter}(1999)}]{Barkai-PhysA1999}
\bibinfo{author}{\bibfnamefont{E.~B.} \bibnamefont{R.~Metzler}}
  \bibnamefont{and} \bibinfo{author}{\bibfnamefont{J.}~\bibnamefont{Klafter}},
  \bibinfo{journal}{Physica A} \textbf{\bibinfo{volume}{266}},
  \bibinfo{pages}{343} (\bibinfo{year}{1999}).

\bibitem[{\citenamefont{Weiss}(1994)}]{Weiss1994}
\bibinfo{author}{\bibfnamefont{G.}~\bibnamefont{Weiss}},
  \emph{\bibinfo{title}{Aspects and {A}pplications of the {R}andom {W}alk}}
  (\bibinfo{publisher}{North-Holland}, \bibinfo{year}{1994}).

\bibitem[{\citenamefont{Majumdar}(2010)}]{Majumdar-PhysA2010}
\bibinfo{author}{\bibfnamefont{S.~N.} \bibnamefont{Majumdar}},
  \bibinfo{journal}{Physica A} \textbf{\bibinfo{volume}{389}},
  \bibinfo{pages}{4299} (\bibinfo{year}{2010}).

\bibitem[{\citenamefont{Redner}(2001)}]{Redner2001}
\bibinfo{author}{\bibfnamefont{S.}~\bibnamefont{Redner}},
  \emph{\bibinfo{title}{A {G}uide to {F}irst-{P}assage {P}rocesses}}
  (\bibinfo{publisher}{Cambridge}, \bibinfo{year}{2001}).

\bibitem[{\citenamefont{R.~Metzler}(2013)}]{Redner2013}
\bibinfo{editor}{\bibfnamefont{S.~R.} \bibnamefont{R.~Metzler},
  \bibfnamefont{G.~Oshanin}}, ed., \emph{\bibinfo{title}{First Passage
  Phenomena and Their Applications}} (\bibinfo{publisher}{World Scientific},
  \bibinfo{year}{2013}).

\bibitem[{\citenamefont{Montroll and Weiss}(1965)}]{Montroll-JMP1965}
\bibinfo{author}{\bibfnamefont{E.}~\bibnamefont{Montroll}} \bibnamefont{and}
  \bibinfo{author}{\bibfnamefont{G.}~\bibnamefont{Weiss}}, \bibinfo{journal}{J.
  Math. Phys.} \textbf{\bibinfo{volume}{6}}, \bibinfo{pages}{167}
  (\bibinfo{year}{1965}).

\bibitem[{\citenamefont{Klafter and Sokolov}(2011)}]{Sokolov2011}
\bibinfo{author}{\bibfnamefont{J.}~\bibnamefont{Klafter}} \bibnamefont{and}
  \bibinfo{author}{\bibfnamefont{I.}~\bibnamefont{Sokolov}},
  \emph{\bibinfo{title}{First Steps in Random Walks}}
  (\bibinfo{publisher}{{O}xford {P}ress}, \bibinfo{year}{2011}).

\end{thebibliography}

\end{document}